\documentclass[twocolumn, showpacs,floatfix]{revtex4}

\usepackage{graphicx}
\usepackage{amsmath}
\usepackage{amssymb}

\newcommand{\bnen}{\begin{equation}}
\newcommand{\eden}{\end{equation}}
\newcommand{\bean}{\begin{eqnarray}}
\newcommand{\eean}{\end{eqnarray}}

\newcommand{\bna}{\begin{array}}
\newcommand{\eda}{\end{array}}

\begin{document}

\title{Periodic Anderson model with $d$-$f$ interaction}

\author{I. Hagym\'asi$^{1,2}$}\email{hagymasi@szfki.hu}
\author{K. Itai$^1$}
\author{J. S\'olyom$^{1}$}

\affiliation{$^1$Research Institute for Solid State Physics and Optics of the 
Hungarian Academy of Sciences, Budapest, H-1525 P.O. Box 49, Hungary\\
$^2$Institute of Physics, E\"otv\"os University, Budapest, P\'azm\'any P\'eter s\'et\'any 1/A, H-1117, Hungary}

\begin{abstract}
We investigate an extended version of the periodic Anderson model where an interaction 
is switched on between the doubly occupied $d$- and $f$-sites. We perform variational 
calculations using the Gutzwiller trial wave function. We calculate the $f$-level 
occupancy as a function of the $f$-level energy with different interaction strengths. 
It is shown that the region of valence transition is sharpened due to the new interaction.

\end{abstract}


\maketitle

\section{Introduction}
The heavy-fermion phenomenon, which is observable in many rare-earth materials is still an 
active field of researched area. The simplest model, which can account for these phenomena 
is the periodic Anderson model. For a review on this topic see \cite{Fazekas:konyv,Hewson:konyv,Fulde:SSPh}. 
The model in its simplest form describes a hybridization ($v$) between a wide conduction band 
and correlated $f$-electrons localized at lattice sites. Despite its simple form there is no general solution to this problem, exact solutions have been obtained only in some special cases \cite{Gulacsi:exact}. Nonperturbative approaches 
like variational methods
\cite{Gutzwiller:original,Vollhardt:RevMod,Rice&Ueda:variational,Fazekas:variational,%
Lamba:variational} are applicable even if the Coulomb interaction ($U_f$) between the 
$f$-electrons is large, though we have to use some uncontrolled approximation 
when calculating the expectation values. In a recent paper \cite{Miyake_cikk} it was shown that 
the repulsion between the $d$- and $f$-electrons plays an important role in valence transition. It has been argued that the sudden change of valence of Ce can account for the presence of a sharp peak in the transition temperature of some superconducting Ce based compounds. The effects of this new interaction have been examined with dynamical mean field theory \cite{Koga:DMFT_Ud&Udf,Saiga:EPAM_DMFT,Hirashima:DMFT_Udf}, slave-boson method \cite{Miyake_cikk,Onishi:EPAM_slbos}, variational Monte Carlo technique \cite{Onishi:EPAM_VMC} and projector-based renormalization approach \cite{Phan:EPAM_renorm}. It has been found in these papers that intermediate valent region narrows in the presence of $d$-$f$ coupling. 

In this paper we examine the ground state properties of this extended model using the Gutzwiller 
method. However, the conventional $d$-$f$ interaction
\begin{equation}
U_{df}\sum_{\boldsymbol{j},\sigma,\sigma'}\hat{n}^f_{\boldsymbol{j\sigma}}
    \hat{n}^d_{\boldsymbol{j\sigma'}}\label{conventional-Udf}
\end{equation}	 
is too difficult to handle within the framework of this method, so we consider a
modified $d$-$f$ interaction
\begin{equation} 
         \widetilde{U}_{df}\sum_{\boldsymbol{j}}\hat{n}^f_{\boldsymbol{j}\uparrow}
    \hat{n}^f_{\boldsymbol{j}\downarrow}\hat{n}^d_{\boldsymbol{j}\uparrow}
   \hat{n}^d_{\boldsymbol{j}\downarrow},
\end{equation} 
where electrons repel each other only when site $\boldsymbol{j}$ is fully occupied. The reason for this modification is that it can be treated easier, since much less electron configurations need to be taken into account. 

\section{Variational calculation}
We summarize the main steps of the performed 
variational calculation following Ref. \cite{Itai:variational}. For
simplicity in the present work we restrict ourselves to a nonmagnetic half-filled band, 
although the variational procedure can be carried out without these assumptions. 
Assuming that there are $N_{\uparrow} = N_{\downarrow}$ up- and down-spin electrons in 
an arbitrary dimensional lattice with $L$ lattice sites, we consider the following 
Hamiltonian:
\begin{gather}
 \mathcal{H}=\sum_{\boldsymbol{k},\sigma}\varepsilon_d(\boldsymbol{k})
 \hat{d}_{\boldsymbol{k},\sigma}^{\dagger}\hat{d}_{\boldsymbol{k},\sigma}
 -v\sum_{\boldsymbol{j},\sigma}\big(\hat{f}_{\boldsymbol{j},\sigma}^{\dagger}
 \hat{d}_{\boldsymbol{j},\sigma}+\hat{d}_{\boldsymbol{j},\sigma}^{\dagger}
 \hat{f}_{\boldsymbol{j},\sigma}\big) \nonumber\\
  +\epsilon_f\sum_{\boldsymbol{j},\sigma}\hat{n}^f_{\boldsymbol{j},\sigma}
 +U_f\sum_{\boldsymbol{j}}\hat{n}^f_{\boldsymbol{j}\uparrow}
 \hat{n}^f_{\boldsymbol{j}\downarrow}+\widetilde{U}_{df}\sum_{\boldsymbol{j}}
 \hat{n}^f_{\boldsymbol{j}\uparrow}\hat{n}^f_{\boldsymbol{j}\downarrow}
 \hat{n}^d_{\boldsymbol{j}\uparrow}\hat{n}^d_{\boldsymbol{j}\downarrow},
\end{gather}
where 
$\hat{n}^f_{\boldsymbol{j},\sigma}=\hat{f}_{\boldsymbol{j},\sigma}^{\dagger}
\hat{f}_{\boldsymbol{j},\sigma}$, and $\hat{n}^d_{\boldsymbol{j},\sigma}=\hat{d}_{\boldsymbol{j},\sigma}^{\dagger}\hat{d}_{\boldsymbol{j},\sigma}$. 
The symbol $\boldsymbol{k}$ denotes the wave vector, and $\boldsymbol{j}$ the site index. 
The width of the $d$-band is $W$. The trial wave function of 
Gutzwiller-type is expressed as
\begin{gather}
 |\Psi\rangle=\hat{P}_{\rm G}^{df}\hat{P}_{\rm G}^f\prod_{\boldsymbol{k}}\prod_{\sigma}[u_{\boldsymbol{k}}
 \hat{f}_{\boldsymbol{k},\sigma}^{\dagger}+v_{\boldsymbol{k}}
 \hat{d}_{\boldsymbol{k},\sigma}^{\dagger}]|0\rangle,
\end{gather}
where the mixing amplitudes $u_{\boldsymbol{k}}/v_{\boldsymbol{k}}$ are also variational parameters, and
$\hat{P}_{\rm G}^{df}$ and $\hat{P}_G^f$ are the Gutzwiller projectors, which are written down as follows:
\begin{eqnarray}
 \hat{P}_{\rm G}^{df} & = & \prod_{\boldsymbol{j}}[1-(1-\eta_{df})\hat{n}^f_{\boldsymbol{j}\uparrow}
 \hat{n}^f_{\boldsymbol{j}\downarrow}\hat{n}^d_{\boldsymbol{j}\uparrow}
 \hat{n}^d_{\boldsymbol{j}\downarrow}],\\
 \hat{P}_{\rm G}^f & = & \prod_{\boldsymbol{j}}[1-(1-\eta_f)\hat{n}_{\boldsymbol{j}\uparrow}^f
 \hat{n}_{\boldsymbol{j}\downarrow}^f],
\end{eqnarray}
where the variational parameters $\eta_{df}$ and $\eta_f$ are controlled by $\widetilde{U}_{df}$ 
and $U_f$, respectively. We use the so-called Gutzwiller approximation to evaluate the expectation values of the Hamiltonian. Optimizing the ground state energy density with respect to the mixing amplitudes, we obtain the following 
expression for that:
\begin{gather}
 \mathcal{E}=\frac{1}{L}\sum_{\boldsymbol{k}\in 
\mathrm{FS}}\left[q_d\epsilon_d(\boldsymbol{k})+\tilde{\epsilon}_f
  -\sqrt{(q_d\epsilon_d(\boldsymbol{k})-\tilde{\epsilon}_f)^2+4\tilde{v}^2}\right] \nonumber\\
     +(\epsilon_f-\tilde{\epsilon}_f)n_f+\widetilde{U}_{df}\nu_{df}+U_f\nu_f,
\label{eq:energy}
\end{gather}
where $n_{f}$ is the average number of $f$--electrons per site, i.e.
\begin{equation}
   n_{f}=\frac{1}{L}\left\langle\sum_{\sigma}\hat{n}^{f}_{\boldsymbol{j},\sigma}\right\rangle,
\end{equation}
$\nu_f$ denotes the density of the doubly occupied $f$-sites and $\nu_{df}$ is the density of such sites when both the $f$- and $d$-sites are doubly occupied, $\tilde{\epsilon}_f$ is the quasiparticle energy level of the $f$-electron (the effective $f$-level, its precise form is not presented here due to its lengthy form), $\tilde{v}$ denotes the renormalized hybridized amplitude, $\tilde{v}$=$\sqrt{q_dq_f}v$, and the summation is carried out over all wave numbers, ${\boldsymbol k}$, of electrons in the Fermi sea (FS).
The $q_d$ and $q_f$ are the kinetic energy renormalization factors for the $d$-electrons and $f$-electrons respectively and are functions of $n$ (the total numbers of electrons per site), $n_f$, $\nu_d$ (the density of the doubly occupied $d$-sites), $\nu_f$, and $\nu_{df}$. The determination of these quantities is the main task of the Gutzwiller method. Their forms in the present model are much more complicated than in those models which do not contain the interaction $\widetilde{U}_{df}$. The results are written in the following complete square (Gutzwiller-like) form:
\begin{widetext}
 \begin{gather}
 q_f=\frac{1}{\frac{n_f}{2}(1-\frac{n_f}{2})}\left(\sqrt{\left(\frac{n_f}{2}
 -\nu_f\right)\left(1-n_f+\nu_f\right)}+\sqrt{\frac{\left(\frac{n_f}{2}
 -\nu_f\right)(\nu_f-\nu_{df})(1-\nu_f-\nu_d)}{1-\nu_f}}
 +\sqrt{\frac{\nu_d\nu_{df}\left(\frac{n_f}{2}-\nu_f\right)}{1-\nu_f}}\right)^2,\\
q_d=\frac{1}{\frac{n-n_f}{2}(1-\frac{n-n_f}{2})}\left(\sqrt{\left(\frac{n-n_f}{2}-
\nu_{df}-\nu_d\right)(1-n+n_f+\nu_d+\nu_{df})}+\sqrt{\frac{\nu_d\left(\frac{n-n_f}{2}-\nu_d-\nu_{df}\right)
(1-\nu_d-\nu_f)}{1-\nu_d-\nu_{df}}} \right.\nonumber\\
\left. + \sqrt{\frac{\nu_{df}(\nu_f-\nu_{df})\left(\frac{n-n_f}{2}-\nu_d-\nu_{df}\right)}
{1-\nu_d-\nu_{df}}}\right)^2, \label{eq:qd_formula}
\end{gather}
\end{widetext}
where $n$ is the total number of electrons per site, i.e. $n=\frac{1}{L}\left\langle\sum_{\boldsymbol{j},\sigma}\hat{n}^{f}_{\boldsymbol{j},\sigma}+\sum_{\boldsymbol{j},\sigma}\hat{n}^{d}_{\boldsymbol{j},\sigma}\right\rangle.$
It is worth mentioning that without $\widetilde{U}_{df}$, $q_f$ depends only on $n_f$ and $\nu_f$ \cite{Itai:variational}.
 Since there is no direct hopping between the different $f$-sites, $q_f$ appears only in the renormalized hybridized amplitude $\tilde{v}$=$\sqrt{q_dq_f}v$. It should be emphasized that $q_d$ is not unity, i.e.\ the width of the $d$-band is narrowed, though there is no Coulomb repulsion between the $d$-electrons. It is also remarkable that the renormalization amplitude of the hybridization still can be written as the square root of the product of $q_f$ and $q_d$ in spite of this new interaction. These are the main results of this paper.

\section{Results}
\begin{figure}[!ht]
\includegraphics[scale=0.5]{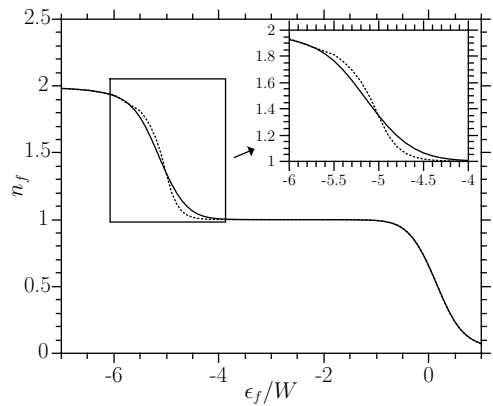}
\caption{\label{nf_ef:fig}
The $f$-level occupancy as a function of the energy $\epsilon_f$ of the $f$-level at 
half-filling, $v/W=0.1875$ and $U_f/W=5$. The continuous and dotted curves belong 
to $\widetilde{U}_{df}/W=0$ and $10$, respectively. The inset shows the enlarged view of the framed region.}
\end{figure}

For further calculations we assume that the density of states of the $d$-band is constant, $\rho(\epsilon)=1/W$, when
$\epsilon$ is in the interval $[-W/2,W/2]$. Using this assumption the 
summation over the wave vectors in Eq. (\ref{eq:energy}) can still be performed analytically.  
The ground state energy density thus obtained should be optimized with respect to $\nu_d$, $\nu_f$, $\nu_{df}$ 
and $n_f$. 
After this procedure we arrive at a nonlinear system of equations for these 
unknown quantities. We solve the obtained equations numerically. Typical results are shown 
in Fig. \ref{nf_ef:fig}. 
One can see that switching on $\widetilde{U}_{df}$ also results in the narrowing 
of the intermediate valent regime of 2$>n_f>$1 as the conventional $d$-$f$ interaction (see Eq.(\ref{conventional-Udf})) \cite{Miyake_cikk,Phan:EPAM_renorm}, while no narrowing occurs in the region of 1$>n_f>$0, since the $d$-$f$ interaction we used is effective only when there are many doubly occupied $f$- and $d$-sites. The conventional $d$-$f$ interaction significantly narrows
the intermediate valent regime of 1$>n_f>$0, too.

\begin{figure}[!ht]
\includegraphics[scale=0.5]{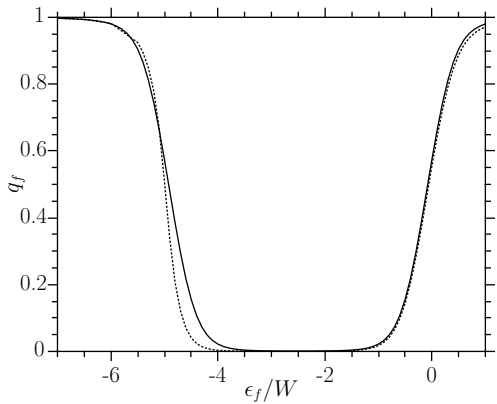}
\caption{\label{qf_ef:fig}
The $f$-level renormalization factor as a function of the energy $\epsilon_f$ of the 
$f$-level at half-filling, $v/W=0.1875$ and $U_f/W=5$. The continuous and dotted curves belong 
to $\widetilde{U}_{df}/W=0$ and $10$, respectively.}
\end{figure}

It is interesting to investigate the $q_f$ renormalization factor, since it is related 
to the effective mass ($m^*\sim q_f^{-1}$). The results are shown in Fig. \ref{qf_ef:fig}. 
The same effect can be seen here too, furthermore the heavy-fermion regime extends to smaller $f$-level energies. 

\begin{figure}[!ht]
\includegraphics[scale=0.5]{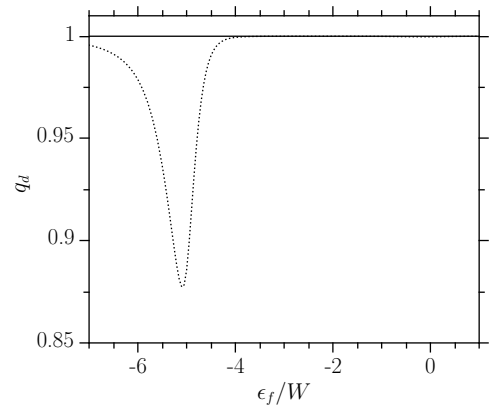}
\caption{\label{qd_ef:fig}
The $d$-band renormalization factor as a function of the energy $\epsilon_f$ of the 
$f$-level at half-filling, $v/W=0.1875$ and $U_f/W=5$. The continuous and dotted curves belong 
to $\widetilde{U}_{df}/W=0$ and $10$, respectively.}
\end{figure}

In Fig. \ref{qd_ef:fig} the renormalization factor for the $d$-band is plotted. As it was outlined in Eq. (\ref{eq:qd_formula}) the $d$-electrons become correlated in the presence of $\widetilde{U}_{df}$, despite the fact that there is no direct $d$-$d$ Coulomb interaction present in the model. 
\section{Conclusions}

We have investigated an extended version of the usual periodic 
Anderson model, where an additional repulsive interaction among the $d$ and $f$ electrons 
was taken into account. We have shown analytically that the conduction band is narrowed owing to the new interaction. We have demonstrated that a finite $\widetilde{U}_{df}$ expands the regime of heavy-fermion character to a lower value of $\epsilon_f$ and narrows the intermediate valent regime at the same time. It is expected that the full $d$-$f$ interaction (Eq. (\ref{conventional-Udf})) has an even sharper effect. The calculation is in progress.

\vspace{4mm}

\noindent \emph{Acknowledgements:} This research was supported by the OTKA grant T68340.

\end{document}